\documentclass[aps,twocolumn,showpacs,superscriptaddress]{revtex4}
\usepackage{graphicx}
\usepackage[normalem]{ulem}
\usepackage{graphicx,color}
\usepackage{amsmath}
\usepackage{amssymb}
\usepackage{amsfonts}
\usepackage{bm}
\def\<{\langle}
\def\>{\rangle}
\newcommand{\ket}[1]{\left|#1\right\rangle}
\newcommand{\bra}[1]{\left\langle#1\right|}
\newcommand{\kpsi}{\left|\psi\right\rangle}
\newtheorem{Theorem}{Theorem}

\newtheorem{Corollary}{Corollary}
\newtheorem{Definition}{Definition}

\begin{document}

\title[]{Interaction-free evolving states of a bipartite system}


\author{ A. Napoli}

\address{Dipartimento di Fisica e Chimica
via Archirafi 36, 90123 Palermo, Italy }

\author{M. Guccione}

\address{Dipartimento di Fisica e Chimica
via Archirafi 36, 90123 Palermo, Italy }

\author{ A. Messina}

\address{Dipartimento di Fisica e Chimica
via Archirafi 36, 90123 Palermo, Italy }

\author{ D. Chru\'sci\'nski}

\address{Institute of Physics, Faculty of Physics, Astronomy and Informatics \\ Nicolaus Copernicus University,  Grudzi{a}dzka 5/7, 87--100 Torun, Poland}

\begin{abstract}
We show that two interacting physical systems may admit entangled
pure or non separable mixed states evolving in time as if the
mutual interaction hamiltonian were absent. In this paper we
define these states Interaction Free Evolving (IFE) states and
characterize their existence for a generic binary system described
by a time independent Hamiltonian. A comparison between IFE subspace
and the decoherence free subspace is reported. The set of all  
pure IFE states is explicitly constructed for a non homogeneous
spin star system model
\end{abstract}


\maketitle

\medskip
\pagebreak

\section{Introduction}

Consider a bipartite system {\textit{S}} consisting of two quantum interacting subsystems \textit{A} and \textit{B} with free Hamiltonians $H_A$ acting on the Hilbert space $\mathcal{H}_A$ and $H_B$ acting on $\mathcal{H}_B$  respectively. The states of \textit{A+B} live in the Hilbert space  $\mathcal{H}=\mathcal{H}_A\otimes\mathcal{H}_B$  where the Hamiltonian of the bipartite system is
\begin{equation}\label{H}
H=H_A+H_B+H_I=H_0+H_I,
\end{equation}
$H_I$ being the operator describing the coupling between
\textit{A} and \textit{B}. Generally speaking, the entanglement
exhibited in the initial pure or mixed state of the bipartite
system, regardless of how it is measured, undergoes changes over
time traceable back to the presence of $H_I$ in the Hamiltonian.
Thus, for example, an initial factorized pure state or a separable
mixed state evolves into an entangled state where, hence,
time-dependent classical and/or quantum correlations between
\textit{A} and \textit{B} generally emerge. In such a general
dynamical scenario it is not surprising the increasing attention
reserved to the existence in some bipartite systems of subradiant
states that is selected pure factorized states which evolve
keeping the system in its fully initial decorrelated condition at
any time instant. Such a peculiar behavior, of both fundamental
\cite{dicke}, \cite{gross} and applicative interest
\cite{petrosyan1}-\cite{ferrari} , results from quantum
interference effects exactly canceling in the evolved state at a
generic time instant right those contributions, stemming from the
superposition principle, which, otherwise, would determine the
onset and possibly the persistence of correlation manifestations
in between \textit{A} and \textit{B}. Subradiance is a cooperative
effect investigated both theoretically \cite{dicke, gross},
\cite{sitek} - \cite{petrosyan2}   and experimentally
\cite{pavolini} - \cite{kalachev1}  after the seminal Dicke paper
\cite{dicke}, mainly in radiation-matter systems where it
describes optically inactive states of atomic ensemble (\textit{A})
in an electromagnetic environment (\textit{B}). The current
upsurge of interest toward these states reflects indeed the
existence of many other physical contexts where this phenomenon
may find promising applications \cite{kalachev0}, \cite{zanardi1}
- \cite{ kalachev2} as well as the experimental evidence that a
system made up of superconducting qubits or a diatomic molecule in
an optical lattice may be prepared in subradiant states. In
connection with such an enlarged view we appropriately remind that
the denomination subradiant states has been adopted
\cite{benivegna3} also to classify factorized states of generic
bipartite systems from which the two subsystems evolve with no
energy exchange between them, maintaining moreover their
statistical independence. In this paper we call subradiant state a generalized state
of this type, that is regardless of the specific nature of both
the subsystems. 

 Recently, for
instance, the environmental noise plaguing the unitary evolution
of superconducting artificial atoms in a circuit QED setting, has
been modeled coupling the dynamical variables of the circuit to
the degrees of freedom of a fermionic bath. Systems of this type,
where bosonic degrees of freedom are absent, might admit
subradiant states under appropriate conditions \cite{ashhab}.

In this paper we go beyond the original notion of subradiance
wondering on the existence of even initially entangled pure or
mixed states of the bipartite system evolving as if \textit{A} and
\textit{B} were decoupled. This condition, guaranting the absence
of energy exchanges between the two subsystems, might be of
interest in any applicative protocol based on quantum processes
involving storage steps. Another dynamical property of such states
is that the quantum covariance of any pair of obsevables
$\mathcal{O}_A$ and $\mathcal{O}_B$ acting on $\mathcal{H}_A$ and
$\mathcal{H}_B$, each one invariant  with respect to the free
 evolution  of the corresponding  subsystem, keeps its initial value
 even if such observables do not commute  with
$H_I$. When states of this kind exist, we call them
interaction-free evolving (IFE) states of the bipartite system.
These states should not be confused with decoherence free states
giving rise to celebrated decoherence free subspaces DFS (see e.g.
review paper \cite{Lidar}). DFS are analyzed in the context of
non-unitary evolution of an open quantum system living in some
Hilbert space $\mathcal{H}$. One says that a linear subspace
$\widetilde{\mathcal{H}} \subset \mathcal{H}$ provides a DFS if
the evolution of the system restricted to
$\widetilde{\mathcal{H}}$ is unitary. Hence if the initial state
vector belongs $\widetilde{\mathcal{H}}$ it stays there and hence
does not lose quantum coherence.  Here, we assume that the
evolution of the bipartite system is unitary on $\mathcal{H}_A
\otimes \mathcal{H}_B$. Of course the subsystem B might be such to
play the role of environment of A. We emphasize that in this case
too an IFE state is a state of the compound system A+B, unitarily
evolving on $\mathcal{H}_A \otimes \mathcal{H}_B$.  Generally
speaking, as previously underlined, in an IFE state A and B
exhibit entanglement at all times even if it might happen as well
that an IFE state keeps a factorized form
$|\psi_A(t)\>|\psi_B(t)\>$  as time progresses. In this case the
state  $|\psi_A(t)\>$, belonging to  $\mathcal{H}_A$, is indeed a
decoherence free state since $|\psi_A(t)\>=\exp(-iH_A t)|\psi_A(0)\>$, by IFE state definition. Recall that if we
consider a non-unitary evolution as a reduction of the unitary one
when the system $S$ is coupled to an environment $E$ and the
interaction system-environment Hamiltonian reads $H_{SE} =
\sum_\alpha S_\alpha \otimes E_\alpha$, then DFS is spanned by
vectors $|\psi\>$ satisfying $S_\alpha |\psi\> = \lambda_\alpha
|\psi\>$ \cite{Lidar}. Hence, if all systems operators
$S_\alpha$ are Hermitian, then a nontrivial DFS
$\widetilde{\mathcal{H}}$ exists only when all $S_\alpha$ mutually
commute on $\widetilde{\mathcal{H}}$. Interestingly, as we show
 in this paper, a similar condition governs the existence of   IFE states.

The main result of this paper is the construction of the
characteristic equation for both pure and mixed  IFE states, that
is the equation whose set of solutions singles out all and only
the IFE states of a given bipartite system. In order to
demonstrate the practical usefulness of such an  equation, we
solve it in the non trivial  case of a non-homogeneous spin star
system finding all its IFE pure states.

\section{IFE pure states}

Let us consider the following

\begin{Definition} \label{DEF-P} A normalized vector  $\ket{\psi} \in \mathcal{H}$ is an IFE pure state if
it satisfies the following equation
\begin{equation}\label{}
  e^{-iHt} \ket{\psi}  \sim e^{-iH_0 t} \ket{\psi}\ ,
\end{equation}
where  `$\sim$'  denotes an equivalence relation: $\ket{\psi} \sim \ket{\phi}$ iff $\ket{\psi} = e^{i\alpha} \ket{\phi}$ with $\alpha$ being a real number (a relative phase).
\end{Definition}
It means that $\ket{\psi}$ is an IFE  state  iff there exists $\alpha \in \mathbb{R}$ such that
\begin{equation}\label{evoluzione con H0}
  e^{-iHt} \ket{\psi}  = e^{-i\alpha t} e^{-iH_0 t} \ket{\psi}\ ,
\end{equation}
at any time instant $t$. In order to characterize all the IFE pure states of the system, let us begin by stating that  $\ket\psi$ is a solution of eq. (\ref{evoluzione con H0}) if, for any nonnegative integer $n$,
\begin{equation}\label{condizione H^n}
{H}^n\ket\psi=({H}_0 + \alpha \mathbb{I})^n\ket\psi
\end{equation}
which implies that eq. (\ref{evoluzione con H0}) is satisfied for
all $t$. For $n=1$ one obtains
\begin{equation}\label{}
  H_I \ket{\psi} = \alpha \ket{\psi} \ ,
\end{equation}
that is, $\ket{\psi}$ defines an eigenvector of $H_I$ and $\alpha$ denotes the corresponding eigenvalue. It means that $\ket\psi$ is a zero-mode of   $H_I^{(\alpha)} := {H}_I - \alpha \mathbb{I}$, i.e.
\begin{equation}\label{condizione n1}
\ket\psi \in {\rm Ker}\, {H}_I^{(\alpha)}\ .
\end{equation}
Moreover, starting from eq. (\ref{condizione H^n}) and exploiting eq. (\ref{condizione n1}) we also obtain
\begin{equation}\label{condizione n2}
{H}_I^{(\alpha)} {H}_0\ket\psi=0
\end{equation}
and, by induction
\begin{equation}\label{CNS}
{H}_I^{(\alpha)} {H}_0^n\ket\psi=0\ ,
\end{equation}
for all $n$. Now, for any eigenvalue $\alpha$ of $H_I$ let us define
\begin{equation}\label{Na}
  \mathcal{N}_\alpha := \bigcap_n {\rm Ker}\, ({H}_I^{(\alpha)} {H}_0^n)\ .
\end{equation}
It is clear that $\mathcal{N}_\alpha$ defines a linear subspace of $\mathcal{H}$. Of course it may happen that $\mathcal{N}_\alpha = \{0\}$. It is easy to show that if  $\ket{\psi} \in \mathcal{N}_\alpha \neq \{0\}$, then equation (\ref{evoluzione con H0}) holds. In this way we have proved

\begin{Theorem} A vector $\ket{\psi} \in \mathcal{H}$ is an IFE state iff $\ket{\psi} \in \mathcal{N}_\alpha \neq \{0\}$ for some eigenvalue $\alpha$ of the interaction part $H_I$.
\end{Theorem}
It is clear that the space $\mathcal{N}$ of IFE states is stratified into mutually orthogonal sectors
\begin{equation}\label{}
  \mathcal{N} = \bigcup_\alpha \mathcal{N}_\alpha\ ,
\end{equation}
with $\mathcal{N}_\alpha \perp \mathcal{N}_\beta$ for $\alpha \neq \beta$.
In particular if $\ket{\psi} \in \mathcal{N}_0$ then
\begin{equation}\label{N0}
  e^{-iHt} \ket{\psi}  =  e^{-iH_0 t} \ket{\psi}\ ,
\end{equation}
at any time instant $t$.

Now, we show that the formula (\ref{Na}) defining $\mathcal{N}_\alpha$ may be considerably simplified. Note that
\begin{equation}\label{}
  [H_0,H_I]\Big|_{\mathcal{N}_0}  = 0 \ .
\end{equation}
Indeed, for any $\ket{\psi} \in \mathcal{N}_0$ one finds $H_0H_I\kpsi - H_IH_0\kpsi=0$. Conversely, if $\kpsi \in {\rm Ker}H_I$ and $[H_0,H_I]\kpsi=0$, then $H_IH_0^n\kpsi=0$ for $n=1,2,\ldots $. To prove this let $\mathcal{M} = {\rm Ker}[H_0,H_I]$ and let
 $\{\ket{e_1},\ldots ,\ket{e_r}\}$ be an orthonormal basis in $\mathcal{M}$ such that
\begin{equation}\label{}
  H_0\Big|_{\mathcal{M}} = \sum_{k=1}^r a_k \ket{e_k}\bra{e_k} \ ,
\end{equation}
and
\begin{equation}\label{}
  H_I\Big|_{\mathcal{M}} = \sum_{k=1}^r b_k \ket{e_k}\bra{e_k} \ ,
\end{equation}
provide spectral decompositions of $H_0$ and $H_I$ restricted to $\mathcal{M}$.
Now, let $\kpsi \in {\rm Ker}H_I$ and $\kpsi \in {\rm Ker}[H_0,H_I]$, that is, we assume that ${\rm Ker} H_I \cap \mathcal{M} \neq \{0\}$.
Suppose that ${\rm Ker} H_I \cap \mathcal{M}$ is spanned by  $\{\ket{e_1},\ldots ,\ket{e_l}\}$ with $l \leq r$, that is,
$H_I\Big|_{\mathcal{M}} = \sum_{k=l+1}^r b_k \ket{e_k}\bra{e_k}$ due to $H_I|e_k\> = 0$ for $k=1,\ldots,l$. One immediately
finds
\begin{equation}\label{}
  H_I H_0^n \kpsi = \sum_{k=l+1}^r a_k^nb_k \ket{e_k}\bra{e_k}\psi\> = 0 \ ,
\end{equation}
due to the fact that $\kpsi =  \sum_{k=1}^l x_k \ket{e_k} \in {\rm Ker} H_I \cap \mathcal{M}$. Hence, $ H_I H_0^n \kpsi=0$ whenever  $H_I \kpsi = 0$ and $[H_0,H_I]\kpsi=0$. In a similar way one shows that $ H^{(\alpha)}_I H_0^n \kpsi=0$ whenever  $H^{(\alpha)}_I \kpsi = 0$ and $[H_0,H^{(\alpha)}_I]\kpsi=0$.

\begin{Corollary} The subspace $\mathcal{N}_0$ may be represented as follows
\begin{equation}\label{C1}
  \mathcal{N}_0 = {\rm Ker}\, H_I \cap {\rm Ker}\, [H_0,H_I]\ ,
\end{equation}
and similarly
\begin{equation}\label{C2}
  \mathcal{N}_\alpha = {\rm Ker}\, H_I^{(\alpha)} \cap {\rm Ker}\, [H_0,H_I^{(\alpha)}]\ ,
\end{equation}
for any eigenvalue $\alpha$ of the interaction part $H_I$.
\end{Corollary}
It is clear that to define $\mathcal{N}_\alpha$ one has to solve
eigenvalues of $H_I$ which might be highly nontrivial. One may ask
a simpler question, namely, how to check whether IFE states do
exist. Combining (\ref{C1}) and (\ref{C2}) one arrives at the
following existence condition:
\begin{Corollary}
A Hamiltonian  $H=H_0 + H_I$ allows for IFE states if and only if $  {\rm Ker}\, [H_0,H_I]$ is nontrivial.
\end{Corollary}

Indeed if $\kpsi$ is an IFE state then there exists $\bar{\alpha}\in\mathbb{R}$, eigenstate of $H_I$, such that $\mathcal{N}_{\bar{\alpha}}$ is not trivial. This existence in turn implies that $\kpsi\in {\rm Ker}[H_0, H_I^{\bar{\alpha}}]={\rm Ker}[H_0,H_I]$. Viceversa if $\mathcal{M}= {\rm Ker}[H_0,H_I]$ is not trivial, $H_0$ and $H_I$ may be simultaneously diagonalized in $\mathcal{M}$ and each common eigenstate is an IFE state since it belongs to $\mathcal{N}_\alpha$ for some $\alpha$. We emphasize that had we put $\alpha=0$ in eq. (\ref{evoluzione con H0}), the existence of IFE states belonging to the restricted set accordingly defined, would not be guaranteed by the condition expressed by corollary 2. The reason is that we cannot be sure to find zero among the eigenvalues of $H_I$ restricted to $\mathcal{M}$.

Suppose now that one deals with a bipartite system in $\mathcal{H}=\mathcal{H}_A \otimes \mathcal{H}_B$ described by
\begin{equation}\label{}
H_0 = H_A  +  H_B \ ,
\end{equation}
and the interaction term $H_I$ (to simplify notation we identify $H_A$ with $H_A \otimes \mathbb{I}_B$ and similarly for $H_B$). Note that the corresponding bipartite IFE states  do exhibit absence of energy exchanges between subsystems $A$ and $B$. Indeed, for any $t$ one finds
\begin{eqnarray*}\label{val medio HA costante}
\mathcal{E}_A(t) &:=&
\bra{\psi}e^{i{H}t} {H}_A e^{-i{H}t}\ket{\psi} \\ & =&\bra{\psi}e^{i{H}_0t}{H}_A  e^{-i{H}_0t}\ket{\psi} =
\bra{\psi}{H}_A \ket{\psi} \ ,
\end{eqnarray*}
and
\begin{eqnarray*}\label{val medio HB costante}
\mathcal{E}_B(t) &:=&
\bra{\psi}e^{i{H}t} {H}_Be^{-i{H}t}\ket{\psi} \\ &= &\bra{\psi}e^{i{H}_0t}  {H}_B e^{-i{H}_0t}\ket{\psi}=
\bra{\psi} {H}_B\ket{\psi}\ ,
\end{eqnarray*}
which shows that energies $\mathcal{E}_A(t)$ and $\mathcal{E}_B(t)$ of two subsystems are conserved. Of course the converse is generally not true. Let us consider for example the time evolution obtained starting from a stationary
state of ${H}$. Under this condition the mean values of
both ${H}_A $ and ${H}_B$, as well as of any time-independent
observable of the system, are obviously stationary but the
eigenstates of ${H}$ do not in general satisfy eq.
(\ref{evoluzione con H0}).

\section{IFE mixed states}

In this section we generalize the notion of IFE for mixed states. Denote by $\mathcal{S}(\mathcal{H})$ the space of density operators living in $\mathcal{H}$ and consider the Hamiltonian dynamics generated by (\ref{H}). One has the following generalization of Definition \ref{DEF-P}

\begin{Definition} \label{DEF-M} A density operator  $\rho \in \mathcal{S}(\mathcal{H})$ is an IFE mixed state if
it satisfies the following equation
\begin{equation}\label{}
  e^{-iHt} \rho e^{iHt} = e^{-iH_0t} \rho e^{iH_0t}\ ,
\end{equation}
at any time instant $t\in \mathbb{R}$.
\end{Definition}
It is clear that if $\rho = |\psi\>\<\psi|$, then the above definition reproduces  Definition \ref{DEF-P}.

Let $|\psi^i_\alpha\>$ denotes an orthonormal basis in $\mathcal{N}_\alpha$, that is,
\begin{equation}\label{}
  H_I |\psi^i_\alpha\> = \alpha |\psi^i_\alpha\> \ ,
\end{equation}
for  $i=1,\ldots,n_\alpha={\rm dim}\, \mathcal{N}_\alpha$. One
 immediately has

\begin{Corollary} A density operators $\rho$ defines an IFE mixed state iff
\begin{equation}
\rho=\sum_{\alpha}\sum_{i,j=1}^{n_\alpha}p_\alpha^{(i,j)}\ket{\psi_\alpha^i}\bra{\psi_\alpha^j}\ ,
\end{equation}
where $p_\alpha^{(i,j)} \geq 0$ and $\sum_{\alpha}\sum_{i,j=1}^{n_\alpha} p_\alpha^{(i,j)} =1$.
\end{Corollary}
Let us observe that any IFE mixed state define a direct sum of positive operators
\begin{equation}
\rho=  \bigoplus_{\alpha}  \rho_\alpha \ ,
\end{equation}
where
\begin{equation}\label{}
  \rho_\alpha = \sum_{i,j=1}^{n_\alpha}p_\alpha^{(i,j)}\ket{\psi_\alpha^i}\bra{\psi_\alpha^j}\ ,
\end{equation}
is supported on $\mathcal{N}_\alpha$. Hence, any IFE pure state belongs to single sector $\mathcal{N}_\alpha$ whereas a genuine IFE mixed state defines a mixture of positive operators supported on all sectors    $\mathcal{N}_\alpha$.

Again, it is clear that if one deals with a bi-partite system  and
if $\rho_{AB}$ is IFE state then
\begin{equation}\label{}
  \mathcal{E}_A(t) = {\rm Tr}( e^{-iHt}\rho_{AB} e^{iHt}\, H_A) = {\rm Tr}( \rho_{AB}\, H_A)\ ,
\end{equation}
and the same for $\mathcal{E}_B(t)$. Hence, there is no energy exchange between subsystems $A$ and $B$ for any IFE mixed state.

\section{IFE pure states of a non-homogeneous spin star system}
Consider a non-homogeneous spin star  system consisting of a  central spin coupled to $N$ mutually not
interacting spins around it. The Hamiltonian describing such a system has the form (\ref{H}) with
\begin{eqnarray}\label{H0 e HI}
{H}_0 =\omega_0\sigma_z+\omega\sum_{i=1}^N{\sigma_z^{(i)}} \ ,
\end{eqnarray}
and
\begin{eqnarray}
{H}_I 
=\sum_{i=1}^N{\gamma_i(\sigma_+\sigma_-^{(i)}+\sigma_-\sigma_+^{(i)})} \ .
\end{eqnarray}
The dynamical variables of the central spin are represented by the
Pauli operators $\sigma_z$, $\sigma_{\pm}\equiv
\frac{1}{2}(\sigma_{x}\pm i\sigma_{y})$ whereas the Pauli
operators describing the $i-$th $(i=1,...,N)$ spin are denoted by
by $\sigma_z^{(i)}$, $\sigma_{\pm}^{(i)}\equiv\frac{1}{2}(
\sigma_{x}^{(i)}\pm i \sigma_{y}^{(i)})$.

Considering this physical system as bipartite and the central spin
as one of the two subsystems, the main aim of this section is
the construction of the set of all IFE pure states associated to the spin star system under scrutiny.  In order to do this let us begin by observing that  a normalized state of our
bipartite system can be always written in the form
$\ket{\Psi}=\ket{-}\ket{\psi_-}+\ket{+}\ket{\psi_+}$ where
$\ket{\pm}$ are the eigenstates of $\sigma_z$ with eigenvalues +1
 and -1 respectively whereas $\ket{\psi_{\pm}}$ belong to the Hilbert
space of the system constituted by the spins $1,...,N$ and satisfying the condition $|\psi_+|^2+|\psi_-|^2=1$.

In view of corollary (1) and corollary 2, we must diagonalize $H_0$ and $ H_I$ within the vectorial space $Ker[H_0,H_I]$ provided $dim(Ker[H_0,H_I])>0$. It is easy to demonstrate that the equation $[H_0,H_I]\kpsi=0 $ may be rewritten as follows
 \begin{eqnarray}\label{A}
&&[H_0,H_I]\kpsi=\\ \nonumber
&&2(\omega_0-\omega)\left[\ket+\sum_{i=1}^N{\gamma_i\sigma_-^{(i)}\ket{\psi_-}}- \ket-\sum_{i=1}^N{\gamma_i\sigma_+^{(i)}\ket{\psi_+}}\right]=0
\end{eqnarray}
which in turn requires the existence of solutions for the two equations
\begin{equation}\label{Hi_pm}
\sum_{i=1}^N{\gamma_i\sigma_{\pm}^{(i)}}\ket{\psi_{\pm}}=0
\end{equation}
 We solve eq. (\ref{Hi_pm}), exploiting the method reported in Ref. \cite{benivegna1}:
  let us  introduce the operators $A_{\pm}$
given by
\begin{equation}\label{A_pm}
A_{\pm}=\exp(\sum_{i=1}^N{g_{\pm}^{(i)}\sigma_z^{(i)}})
\end{equation}
where the  complex parameters $g_{\pm}^{(i)}$ will be
chosen later.

The two operators $A_{+}$ and $A_{-}$ thus defined are in general
neither unitary nor Hermitian. However they are not singular and
thus $A_{\pm}^{-1}$ there exist. Accordingly eq.  (\ref{Hi_pm}) may be transformed as follows

\begin{equation}\label{Hi_pm_A}
A_{\pm}^{-1}\sum_{i=1}^2{\gamma_i\sigma_{\pm}^{(i)}}A_{\pm}A_{\pm}^{-1}\ket{\psi_{\pm}}=0
\end{equation}
On the other hand, it is easy to demonstrate that
\begin{equation}\label{A-1sigmaA}
A_{\pm}^{-1}\sigma_{\pm}^{(i)}A_{\pm}=\sigma_{\pm}^{(i)}e^{\mp 2
g_{\pm}^{(i)}}
\end{equation}
and then, choosing the parameters $g_{\pm}^{(i)}$ $(i=1,...,N)$ in such a way
that $\gamma_i=\gamma e^{\pm 2g_{\pm}^{(i)}}$ with
$\gamma=\sqrt{\sum_{i=1}^N{\gamma_i^2}}$, the condition under
which the state
$\ket{\Psi}=\ket{-}\ket{\psi_-}+\ket{+}\ket{\psi_+}$ belongs to
the kernel of $[H_0,H_I]$ becomes
\begin{equation}\label{eq_stati momento angolare}
\sum_{i=1}^N{\gamma\sigma_{+}^{(i)}}(A_{+}^{-1}\ket{\psi_{+}})=0\;\;\;\;
\mathrm{and}
\;\;\sum_{i=1}^N{\gamma\sigma_{-}^{(i)}}(A_{-}^{-1}\ket{\psi_{-}})=0
\end{equation}
These equations show that due to the operators $A_\pm$
 we get rid of the non homogeneous character of Eq. (\ref{Hi_pm_A})  where it appears through the $i$--dependence of the coupling constants $(\gamma_i)$.

Let us note that the choice of the parameters $g_{\pm}^{(i)}$
guarantees that  the two operators $A_+$ and $A_-$ satisfy
$A_+A_-=A_-A_+=I$. Let's moreover observe that the states
$\ket{\widetilde{\psi}_{\pm}}=A_{\pm}^{-1}\ket{\psi_{\pm}}$
satisfying eq. (\ref{eq_stati momento angolare}) are well known in
terms of the simultaneous eigenstates $\ket{r,m,\nu}$ of the
square and of the $z$-component of the total angular momentum of
the $N$ uncoupled spins
\begin{equation}\label{autostati s^2}
S^2\ket{r,m,\nu}\equiv\frac{1}{2}(\sum_{i=1}^{N}\vec{\sigma}^{(i)})^2\ket{r,m,\nu}=r(r+1)\ket{r,m,\nu}
\end{equation}
where $r=0,1,....,\frac{N}{2}$ if $N$ is even and
$r=\frac{1}{2}, \frac{3}{2},...,\frac{N}{2}$ if $N$ is odd. Moreover
\begin{equation}\label{autostati s_z}
S_z\ket{r,m,\nu}\equiv\frac{1}{2}\sum_{i=1}^{N}\sigma_z^{(i)}\ket{r,m,\nu}=m\ket{r,m,\nu}
\end{equation}
with $m=-r-r+1,....,r$. The quantum number $\nu=1,2,....,\nu(r)$ with
\begin{eqnarray}\label{ni(r)}
\nu(r)=\left(\begin{array}{c}
N \\
\frac{N}{2}-r
\end{array}\right) +\left(\begin{array}{c}
N \\
\frac{N}{2}-r-1
\end{array}\right)
\end{eqnarray}
and  $
\left(\begin{array}{c}
N \\
-1
\end{array}\right) =0 $
allows to distinguish between different states of the coupled
angular momentum basis characterized by the same $r$ and $m$.
It is possible to convince oneself that
$\ket{\widetilde{\psi}_{+}}\equiv\sum_{r,
\nu}{C_{r,\nu}^+\ket{r,r,\nu}}$ and
$\ket{\widetilde{\psi}_{-}}\equiv\sum_{r,\nu}{C_{r,\nu}^-\ket{r,-r,\nu}}$
with $C_{r,\nu}^\pm \in\mathbb{C}$. We may thus claim that a
generic state $\kpsi$ satisfying eq. (\ref{A}) may be written as
follows
\begin{equation}\label{sol_A}
\kpsi=\ket{+}\sum_{r, \nu}{C_{r,\nu}^+A_+\ket{r,r,\nu}}+\ket{-}\sum_{r,\nu}{C_{r,\nu}^-A_-\ket{r,-r,\nu}}
\end{equation}
It is remarkable that $Ker[H_0,H_I]$ for the Hamiltonian model
under scrutiny coincides with $Ker{H_I}$ which means that
$H_I\kpsi=0$ iff $\kpsi$ is given by eq. (\ref{sol_A}).  This
result is a direct consequence of the fact that the resolution of
the equation $H_I\kpsi=0$ leads exactly to eqs. (\ref{eq_stati
momento angolare}). In view of corollary 2 we may thus claim that
$\mathcal{N}_\alpha$ is empty for each eigenvalue  $\alpha\neq 0$
of $H_I$. We thus may conclude that the space $\mathcal{N}$ of
IFE pure states for our Hamiltonian model coincides with
$\mathcal{N}_0$. It is interesting to investigate the
diagonalization problem of $H_0$ within
$\mathcal{N}\equiv\mathcal{N}_0$. To this end 
let's observe that both the operators $A_+$ and $A_-$ commute
with the $z$ component of the total angular momentum operator
$S_z$ of the $N$ spins. This property directly implies that
the states $A_+\ket{r,r,\nu}$ as well as the states $A_-\ket{r,-r,\nu}$
are eigenstates of $S_z$ with eigenvalues $r$ and $-r$
respectively. We have indeed
\begin{eqnarray}\label{Apm_autostati s_z}
&&S_z A_{\pm}\ket{r,\pm r,\nu} = A_{\pm}A_{\pm}^{-1}S_zA_{\pm}\ket{r,\pm
r,\nu}\nonumber \\&& \equiv A_{\pm}S_z\ket{r,\pm r,\nu} = \pm rA_{\pm}\ket{r,\pm r,\nu}
\end{eqnarray}

On the other hand, it is immediate to convince oneself that they
are also eigenstates of ${H}_0$ correspondent to the eigenvalues
$(\omega_0+ 2r\omega)$ and $-(\omega_0+ 2r\omega)$ respectively.
This circumstance in turn means that these states are also
eigenstates of the total hamiltonian given by eq. (\ref{H}) being
simultaneous eigenstates of ${H}_0$ and ${H}_I$. In other
words the IFE states space may be represented as a direct sum of
appropriate vectorial subspaces invariant under the action of the
total Hamiltonian $H$. As a consequence we might envision initial
conditions starting from which the system effectively evolves
conserving the value of its initial entanglement no matter  the
measure  used. Our results on the structure of $\mathcal{N}_0$
play an important role in the context of the problem of the
diagonalization of non-homogeneous spin star system hamiltonian
model under scrutiny in this section. In the near past, indeed,
many efforts have been made in order to find the spectrum of such
hamiltonian but, until now only a particular set of eigensolutions
are known \cite{jivulescu}.

\section{Conclusive remarks}

In this paper we have introduced a new class of states of a
bipartite system christened IFE states. This set of states
encompasses all those initial conditions of the compound system
from where each subsystem evolves with no energy exchange with the
other one and leaving unmodified the level of mutual entanglement
whatever measure is adopted. These properties stem from
cooperative effects leading through quantum interference
processes, to the cancellation of any dynamical consequence of the
coupling term $H_I$.We stress that since the constructions of the
IFE states space requires the resolution of their characteristic
equations in the Hilbert state of the given bipartite system, it
may happen that it is empty. It is however worth noticing that
when subradiant states exist then they are IFE states too,
allowing us to claim that our definition of IFE states generalizes
indeed that of subradiant state. Our main result is constituted by
the two characteristic equations of the states (Theorem 1 and
Corollary 3) as well as construction of the set of all the IFE
states of a nontrivial hamiltonian model of evergreen interest. A
remarkable merit of such a result is its universality with respect
to time-independent Hamiltonian models which means that the
characteristic equations here reported are applicable to any
bipartite system  evolving unitarily. The more intriguing situation
corresponding to the evolution of a bipartite system in presence
of an environment is currently under investigation and will be
presented elsewhere.

\section{Acknowledgements}
DC was partially supported by the National Science Center project  
DEC-2011/03/B/ST2/00136.

\end{document}